\documentclass[aps,prb,twocolumn,superscriptaddress,showpacs]{revtex4}
\usepackage{graphicx}

\begin{document}

\title{\textit{B-T} phase diagram of CoCr$_{2}$O$_{4}$ in magnetic fields up to 14 T}

\author{A. V. Pronin}\email{a.pronin@hzdr.de} \author{M. Uhlarz} \author{R. Beyer} \author{T. Fischer} \author{J. Wosnitza}

\address{Dresden High Magnetic Field Laboratory (HLD), Helmholtz-Zentrum Dresden-Rossendorf, 01314 Dresden,
Germany}

\author{B. P. Gorshunov}

\address{A. M. Prokhorov Institute of General Physics, Russian Academy of Sciences,
119991 Moscow, Russia}

\address{Moscow Institute of Physics and Technology (State University), 141700 Dolgoprudny, Moscow Region, Russia}

\address{1. Physikalisches Institut, Universit\"{a}t Stuttgart, Pfaffenwaldring 57, 70550 Stuttgart, Germany}

\author{G. A. Komandin}

\address{A. M. Prokhorov Institute of General Physics, Russian Academy of Sciences,
119991 Moscow, Russia}

\author{A. S. Prokhorov}

\address{A. M. Prokhorov Institute of General Physics, Russian Academy of Sciences,
119991 Moscow, Russia}

\address{Moscow Institute of Physics and Technology (State University), 141700 Dolgoprudny, Moscow Region, Russia}

\author{M. Dressel}

\address{1. Physikalisches Institut, Universit\"{a}t Stuttgart, Pfaffenwaldring 57, 70550 Stuttgart, Germany}

\author{A. A. Bush}

\address{Moscow Institute of Radiotechnics, Electronics, and Automation, 117464 Moscow, Russia}

\author{V. I. Torgashev}

\address{Faculty of Physics, Southern Federal University, 344090
Rostov-on-Don, Russia}

\date{\today}

\begin{abstract}
We have measured the magnetization and specific heat of multiferroic
CoCr$_{2}$O$_{4}$ in magnetic fields up to 14 T. The high-field
magnetization measurements indicate a new phase transition at
$T^{\ast} = 5 - 6$ K. The phase between $T^{\ast}$ and the lock-in
transition at 15 K is characterized by magnetic irreversibility. At
higher magnetic fields, the irreversibility increases. Specific-heat
measurements confirm the transition at $T^{\ast}$, and also show
irreversible behavior. We construct a field-temperature phase
diagram of CoCr$_{2}$O$_{4}$.

\end{abstract}


\maketitle

CoCr$_{2}$O$_{4}$ is a ferrimagnetic spinel with Curie temperature
$T_{C} = 94$ K. \cite{tomiyasu} At $T_{S} \approx 26$ K, a
structural transition happens, below which a short-range-ordered
spiral component develops in the spin system, resulting in a conical
magnetic structure. \cite{menyuk} The structural transition is also
accompanied by the emergence of a spontaneous electric polarization,
which direction can be reversed by applying a magnetic field.
\cite{yamasaki}  At $T_{\rm lock-in} = 15$ K, the period of the spin
spiral becomes commensurate (``locks") to the lattice parameter.
\cite{tomiyasu}

\begin{figure}[]
\centering
\includegraphics[width=\columnwidth,clip]{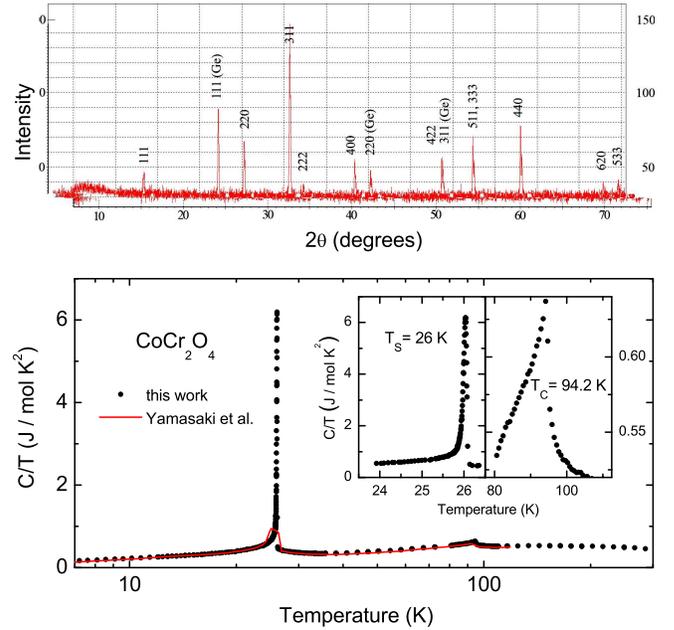}
\vspace{-0.5cm} \caption{(Color online) Top panel: x-ray diffraction
pattern of CoCr$_{2}$O$_{4}$ powder (peaks from a Ge reference are
also marked). Bottom panel: specific heat of polycrystalline
CoCr$_{2}$O$_{4}$. The line represents published single-crystal data
(Ref. \onlinecite{yamasaki}). Insets: close-ups of the structural
($T_{S}$) and ferrimagnetic ($T_{C}$) transitions (note different
vertical scales).} \label{Specific_heat_x_ray}
\end{figure}

CoCr$_{2}$O$_{4}$ is believed to be the first example of a
multiferroic, where both spontaneous magnetization and spontaneous
polarization are of spin origin (\textit{cf.} the spin-current model
of Katsura \textit{et al.}, Ref. \onlinecite{katsura}). As
multiferroics are appealing because of basic physical interest as
well as their potential technological applications, \cite{kimura}
the reported multiferroicity of CoCr$_{2}$O$_{4}$ has triggered a
broad experimental research of the compound. \cite{lawes, choi,
mufti, chang} So far the magnetic and thermal properties of
CoCr$_{2}$O$_{4}$ have been investigated either in zero magnetic
field or in relatively low fields. As the system is very rich for
different types of phase transitions, measurements in high magnetic
fields can shed more light on the nature of the different phases.

Here, we report results of a high-magnetic-field investigation (up
to 14 T) of CoCr$_{2}$O$_{4}$. In the high-field magnetization data,
we have found signs of a new transition at around $T^{\ast} = 5$ K.
The phase between $T^{\ast}$ and $T_{\rm lock-in}$ is characterized
by magnetic irreversibility. Specific-heat measurements confirm the
transition at $T^{\ast}$, and also show the irreversible behavior.
We propose a field-temperature phase diagram of CoCr$_{2}$O$_{4}$.

\begin{figure*}[]
\centering
\includegraphics[width=14.5cm]{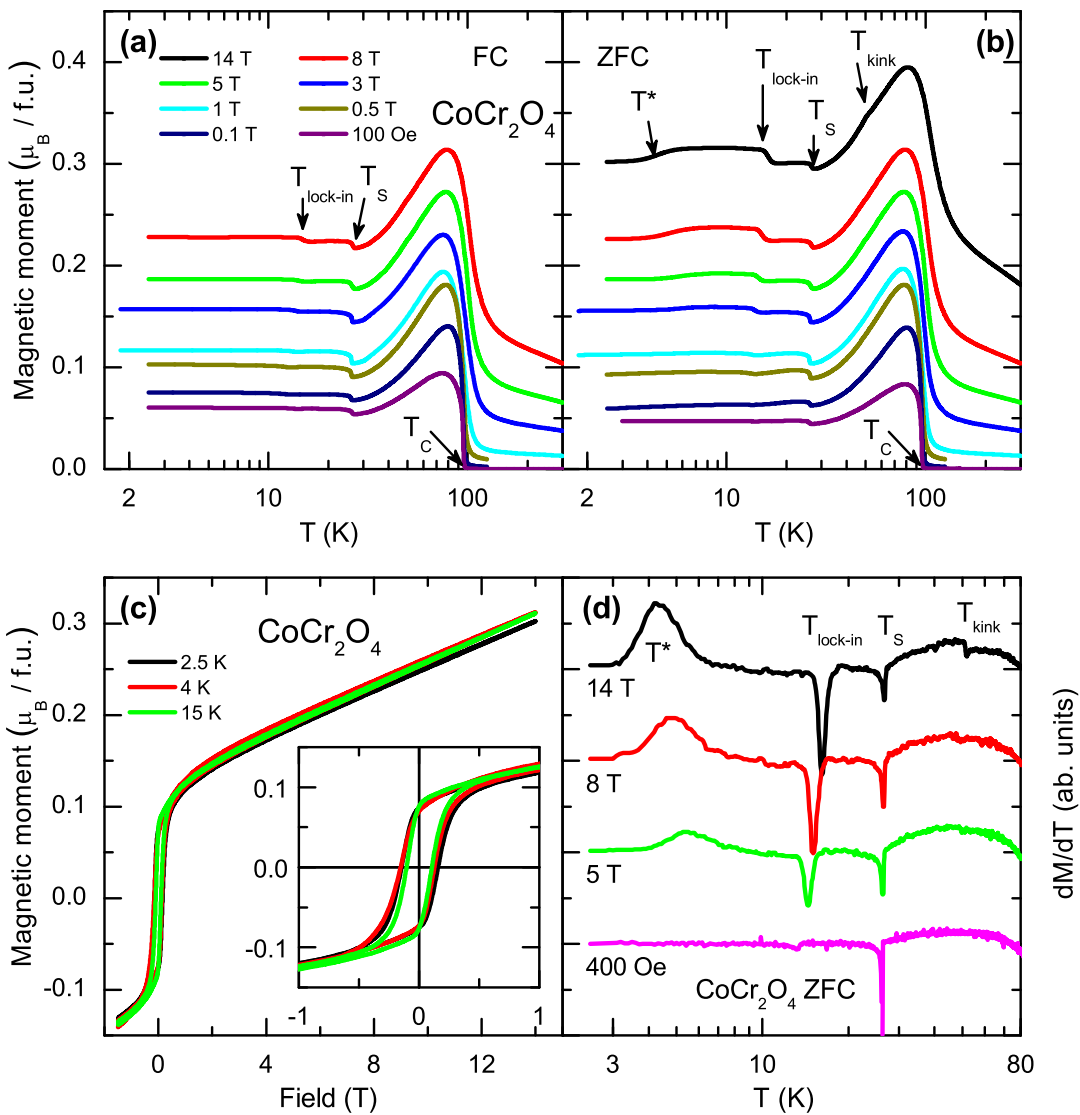}
\vspace{0.0cm} \caption{(Color online) Magnetization of
CoCr$_{2}$O$_{4}$, measured as a function of temperature [(a) - ZFC,
(b) - FC measurements] and magnetic field (c). (d) -- temperature
derivative of the ZFC magnetization, $dM_{\rm ZFC}/dT$, as a
function of temperature. The transition temperatures are marked and
discussed in the text. The inset in (c) shows magnified hysteresis
loops at low fields.} \label{Moment}
\end{figure*}

The CoCr$_{2}$O$_{4}$ samples have been synthesized from
Co$_{3}$O$_{4}$ and Cr$_{2}$O$_{3}$ powders at $T = 1400 ^{\circ}$C.
CoCr$_{2}$O$_{4}$ powder has been pressed into pellets of 10 mm
diameter and 1 -- 2 mm thickness. X-ray diffraction measurements
have been performed at room temperature using a commercial
diffractometer with a position-sensitive detector utilizing
Cu-K$_{\alpha}$ radiation (Fig. \ref{Specific_heat_x_ray}, top
panel). The data have been analyzed by standard Rietveld refinement.
The diffraction pattern shows the cubic symmetry (space group
$Fd\bar{3}m$) with no indication of spurious phases. The lattice
constant, $a = 8.328(2)$ {\AA}, is in good agreement with published
results. \cite{mansour, casado}

Magnetization measurements up to 14 T have been carried out in
commercial SQUID and vibrating-sample-magnetometry setups at
temperatures between 1.8 and 300 K. Specific-heat measurements have
been performed between 2 and 300 K in fields up to 14 T in a
commercial $^{4}$He cryostat. At temperatures below 10 K, we also
used a modified relaxation-calorimetry technique. \cite{wang} For
both, magnetization and specific-heat measurements, the sample was
heated up to room temperature before each temperature sweep in order
to avoid any influence of the sample history on our measurements.

To confirm the quality of our samples, specific-heat measurements
have also been made in zero magnetic field. The outcome of these
measurements is shown in the bottom panel of Fig.
\ref{Specific_heat_x_ray}. At $T_{C} = 94.2$ K and $T_{S} = 26$ K,
the specific heat demonstrates sharp anomalies, associated with the
phase transitions. The first-order nature of the structural
transition is evident. For comparison, data from single-crystal
measurements of Ref. \onlinecite{yamasaki} are shown as well.

In Fig. \ref{Moment} we show the results of temperature-dependent
[(a) and (b)] and field-dependent [(c)] magnetization measurements.
Above 2 T, the magnetization increases linearly with applied field.
Hysteresis loops are shown in the inset of panel (c). The
temperature-dependent data have been collected according to the
standard zero-field-cooled (ZFC) and field-cooled (FC) measurement
protocols. In order to show the low-temperature phase transitions
more vividly, we plot the temperature derivative of the ZFC
magnetization, $dM_{\rm ZFC}/dT$, as a function of temperature in
panel (d).

The FC data well reproduce the published magnetization data on
polycrystalline and single-crystal samples. \cite{tomiyasu, mufti}
The ferrimagnetic ($T_{C}$), structural ($T_{S}$), and lock-in
transitions ($T_{\rm lock-in}$) are clearly visible. Another anomaly
is a kink around 50 K (marked as $T_{\rm kink}$ in Fig.
\ref{Moment}). This kink is most pronounced in the highest field
curve (14 T). We believe the anomaly we observe at $T_{\rm kink}$
should be related to the dielectric-constant anomaly reported by
Lawes \textit{et al.} at around 50 K, and attributed to the
development of short-range spiral magnetic order. \cite{lawes}

At low fields, the ZFC data deviate from the FC curves below the
Curie temperature $T_{C}$. This is in line with observations by
Tomiyasu \textit{et al.}, \cite{tomiyasu} who have interpreted this
irreversibility as an indication of a possible spin-glass state. At
higher fields (1 T and above), this irreversibility diminishes,
which is consistent with the spin-glass picture.

\begin{figure}[t]
\centering
\includegraphics[width=\columnwidth,clip]{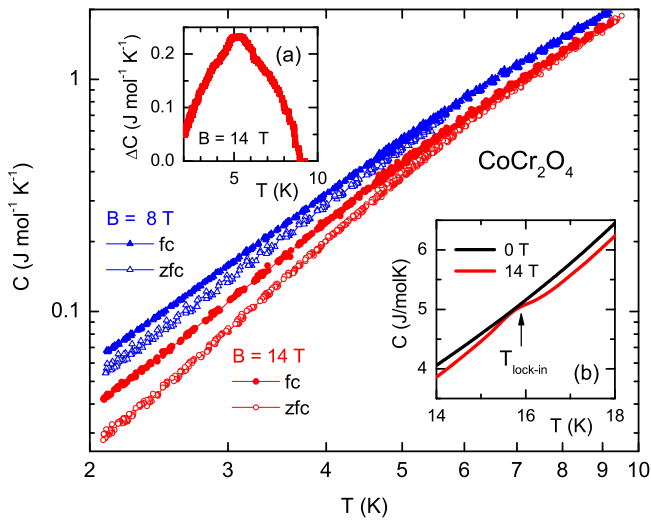}
\vspace{-0.5cm} \caption{(Color online) Specific heat of
CoCr$_{2}$O$_{4}$ at 8 and 14 T, measured as a function of
temperature in FC and ZFC modes. Insets: (a) -- difference between
the FC and ZFC specific heat, $\Delta C = C_{\rm FC} - C_{\rm ZFC}$,
as a function of temperature at 14 T; (b) -- specific heat in 0 and
14 T near the lock-in transition.} \label{Specific_heat}
\end{figure}

In addition to the known transitions, a new transition appears at
$T^{\ast}$ = $5 - 6$ K. This feature is most pronounced in fields
above a few Tesla. In the ZFC magnetization curves a ``bump" appears
at temperatures between $T^{\ast}$ and $T_{\rm lock-in}$. Below
$T^{\ast}$ and above $T_{\rm lock-in}$, the ZFC and FC curves merge.
Remarkably, the bump becomes more pronounced as field increases.
This is quite unusual and is in obvious contradiction with the
simple spin-glass picture, where higher fields would suppress
irreversibility.

\begin{figure}[t]
\centering
\includegraphics[width=\columnwidth,clip]{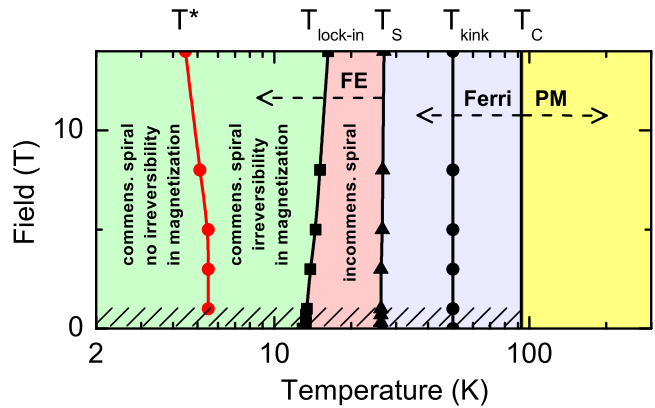}
\vspace{-0.5cm} \caption{(Color online) Field-temperature phase
diagram of CoCr$_{2}$O$_{4}$ based on our magnetization and
specific-heat measurements and on literature data. $T^{\ast}$,
$T_{\rm lock-in}$, $T_{S}$, and $T_{\rm kink}$ are taken as
inflection points of the magnetization \textit{vs.} temperature
curves (Fig. \ref{Moment}). At H = 0, $T_{\rm kink}$ is set to 50 K,
according to Ref. \onlinecite{lawes}. $T_{C}$ is taken equal to its
zero-field value. The transition at $T^{\ast}$ is not visible in the
low-field ($B < 1$ T) data. Possible spin-glass state (Ref.
\onlinecite{tomiyasu}) at low fields is shaded. ``PM", ``Ferri", and
``FE" stand for paramagnetic, ferrimagnetic, and ferroelectric
phases, correspondingly.} \label{Phase_diagram}
\end{figure}

Our specific-heat measurements confirm the transition at $T^{\ast}$,
as well as the irreversible behavior related to this anomaly. In
Fig. \ref{Specific_heat} we show the specific heat $C$, measured at
temperatures around $T^{\ast}$ in ZFC and FC modes at 8 and 14 T.
Overall, the specific heat decreases with increasing magnetic field,
as the magnetic entropy at low temperature decreases in higher
fields. As visible in the double-logarithmic scale in Fig.
\ref{Specific_heat}, at around $T^{\ast}$ the slope of all $C (T)$
curves changes. Most remarkably, around this temperature the ZFC
curves deviate from the FC measurements. The FC specific heat is
always larger than the ZFC data. The deviation, $\Delta C = C_{\rm
FC} - C_{\rm ZFC}$, increases with applied field. This fact is in
line with our magnetization measurements: the irreversibility in
magnetization increases with field.

As temperature approaches zero, $\Delta C$ vanishes (inset (a) of
Fig. \ref{Specific_heat}). Note, that the apparent divergence
between the $C_{\rm FC}$ and $C_{\rm ZFC}$ curves, seen in the main
frame of the figure at $T \rightarrow 0$, is an artifact due to the
double-logarithmic scale.

This new transition at $T^{\ast}$ should be related to some changes
in the spiral magnetic structure. Neutron-scattering measurements in
magnetic fields should be able to clarify this issue.

We would like to note, that one should not expect the transition at
$T^{\ast}$ to be visible as a sharp anomaly in the specific heat.
Even a well-established (and sharper) lock-in transition at 15 -- 16
K, is not seen in zero-field specific heat. Only in 14 T, a small
peak appears (see inset (b) of Fig. \ref{Specific_heat}).

Finally, based on the magnetization and specific-heat measurements,
we plot a field-temperature phase diagram of CoCr$_{2}$O$_{4}$ (Fig.
\ref{Phase_diagram}). As transition temperatures, we take the
inflection points of the magnetization \textit{vs.} temperature
curves (except for $T_{C}$, which is taken equal to its zero-field
value). In addition to the curves shown in Fig. \ref{Moment}, data
collected at 0.3 and 0.7 T have been used for the phase diagram.

Summarizing, we have investigated thermodynamic properties of
CoCr$_{2}$O$_{4}$ in magnetic fields up to 14 T. We have found signs
of a new possible phase transition at $T^{\ast} = 5 - 6$ K. This
transition is associated with an irreversibility in the magnetic
system, as confirmed by magnetization and specific-heat
measurements. Remarkably, the magnetic field stimulates rather than
suppresses the irreversibility effects. We believe, this can be
related to the appearance of the spiral magnetic structure.
Neutron-scattering measurements at high magnetic fields might give
valuable information in clarifying the mechanisms of this
transition.

We thank F. Wolff-Fabris for useful discussions and A. Wobst for
technical assistance. Part of this work was supported by EuroMagNET
II (EU contract No. 228043) and by the Russian Foundation for Basic
Research (Grant No. 09-02-00280-a).

\end{document}